# Mineralization, Grain Growth and Disk Structure: Observations of the Evolution of Dust in Protoplanetary Disks

Dan M. Watson

*Department of Physics and Astronomy, University of Rochester,
Rochester, New York 14627 USA*

**Abstract.** During the past five years, the Spitzer Space Telescope and improved ground-based facilities have enabled a huge increase in the number of circumstellar disks, around young stars of Solar mass or smaller, in which the composition of the solid component has been studied with complete mid-infrared spectra. With these samples we can assess observationally the evolution of dust through the planet-forming era, in parallel with the evolution of the composition and structure of protoplanetary disks. Here we will review the progress in this endeavour, with emphasis on objects in nearby associations and star-formation regions, and on the methods by which dust composition is determined from the infrared spectra of young stellar objects.

## 1. Introduction

Theories of the evolution of disks around young stars, and of planetary formation, confront many decisive new constraints in the era of the Spitzer Space Telescope and 8-10-meter-class ground-based telescopes. The frequency of infrared excesses – and thus the fraction of a cluster of young stellar objects that have dusty circumstellar disks – drops to half by age 3-4 Myr (Haisch et al. 2001; Sicilia-Aguilar et al. 2006; Hernández et al. 2008). Similar timescales are found for accretion onto the central stars (e.g. Muzerolle et al. 2003), and for the frequency of possession of gas in the disk (Najita and Williams 2005). Such results have lent support to theories of photoevaporation (e.g. Alexander et al. 2006), which can dissipate the gas and small dust grains disks on $10^5$ year timescales, starting when accretion onto the star drops below about $10^{-10} M_\odot$ year$^{-1}$. There have also been hints that we have observed the giant-planetary formation time scale: detection of gaps and central clearings in disks, and recent studies of Saturn's moon Iapetus, indicate that giant planets can form at age 1-3 Myr (Calvet et al. 2005; Castillo-Rogez et al. 2007). All of these observational threads have been influential in recent increases in the speed with which core-accretion models can build giant planets (e.g. Ida & Lin 2008; Mordasini et al. 2008).

    What of the evolution of the dust grains themselves? It is also within our grasp to study the composition of grains throughout the era of planet formation. The mid-infrared spectrum of dust carries information on particle sizes and mineral content, which can in principle help us assess the pace and mechanisms of particle growth in early stages of protoplanetary disk development, a crucial ingredient in studies of planetary formation. Such results can also allow us to tie the chronology of protoplanetary disks with that of the Solar nebula. So far, most dust evolutionary timescales

are known only theoretically and under assumption of steady processes. Dust should settle to the midplane (sediment) and grow to $\gg 10$ μm size in $10^{4\text{-}5}$ years (e.g. Goldreich & Ward 1973, Weidenschilling 1997, Dullemond and Dominik 2005). Dust in the inner disk should "mineralize" substantially on similar time scales (e.g. Gail 2004; Harker & Desch 2002). These processes all have time scales much shorter than the duration of the infrared-excess phase; what role do they play in disk evolution?

The following progress report is based mostly upon the first several large, complete surveys of the spectra of silicate dust in clusters of Class II YSOs[1], primarily with the NASA Spitzer Space Telescope and its Infrared Spectrograph (IRS; Houck et al. 2004). The ages of the clusters surveyed to date already span an interesting range of infrared-excess evolution and possible giant-planetary formation.

## 2. Dust-Grain Mineralogy

### 2.1. The Tools: Opacities and Models

Characterization of the grains in a disk means determining the masses and mass fractions of the various sizes and compositions of silicate grains. This is done best by fitting spectra with emissivities derived from empirical optical constants, for grains of given size and shape. The most widely-used empirical opacities are those in the Heidelberg-Jena-St. Petersburg database[2] (Henning et al. 1999, Jäger et al. 2003), a compendium of results now involving several laboratories and scientists from all over Europe, and including measurements on components of primitive meteorites (e.g. Posch et al 2007) as well as natural and synthetic astrophysical analogues. Members of the Earth and Planetary Materials group at Osaka University (e.g. Chihara, Koike, Tsuchiyama) have also produced many measurements of optical properties of astrophysically-interesting minerals. Such research is reviewed extensively elsewhere in this volume.

In young (< 5 Myr) disks the dust is opaque at mid-infrared wavelengths and shortward, in all directions. However, illumination by starlight usually dominates over accretion power in the heating of the parts of the disk that are brightest in the mid-IR. Such heating from outside leads to higher temperatures among dust grains in the first optical depth than is found nearer the disk midplane. The uppermost small dust grains are "superheated" further above the temperature of grains at midplane due to the tendency of such grains to absorb more efficiently at visible and near-infrared wavelengths than they can radiate at mid-IR wavelengths. Throughout protoplanetary disks, the gas density is large enough that dust and gas are in local thermodynamic equilibrium (e.g. Kamp & Dullemond 2004). Thus the spectral features of the uppermost silicate grains are seen in emission. The emitting region can be thought of as a distinct, optically-thin skin lying atop a cooler, opaque background. The temperature and density of such a layer will not vary importantly by much more than does the wavelength we use to probe it: a plane-parallel slab without very many components may serve accurately to extract the relative abundances of the various dust species. These assumptions could lead to significant errors in the dust-species mass fractions in highly-settled, inner disks at high (near edge-on) inclination, but not at low inclination. Since highly-sedimented, highly-inclined disks comprise a small minority of the objects, such errors should not distort substantially the trends we analyze here.

---

[1] Here the Class II young stellar object (YSO) designation is taken to imply solar-type or smaller stars in the T Tauri phase, with large infrared excesses that indicate a surrounding, optically-thick accretion disk, but no remaining trace of the natal envelope.
[2] http://www.mpia-hd.mpg.de/HJPDOC/

Two basic recipes for the spectral decomposition of dust-feature emission from disks are in common use. Both are two-temperature models fit to spectra by minimization of $\chi^2$. They differ in the number of grain sizes used to approximate the real, continuous grain-size distribution, the manner in which grain shape and structure is included, and the spatial distribution of the two temperature components. Despite the differences they seem overall to give similar results when modeling the same spectrum. For simplicity we will refer to them in the following as "East" and "West."

*East* Used for example by van Boekel et al. (2005), Honda et al. (2006), Apai et al. (2005), Sicilia-Aguilar et al. (2007) and Bouwman et al. (2008), to model spectra in the 8-13 μm atmospheric window, or in the broader ranges observed by *ISO* and *Spitzer*. Here the infrared-excess flux density $F_\nu$ is decomposed into two components with different temperatures, one for the continuum ($T_c$) and one for the optically-thin suspended dust grains ($T_w$), which normally correspond to emission from grains deep beneath the disk surface, or nearer to it, respectively:

$$F_\nu = a_0 B_\nu(T_c) + B_\nu(T_w) \sum_i a_i \kappa_{\nu i} + a_{PAH} I_\nu^{PAH} \quad , \tag{1}$$

where the *a*s are adjustable parameters whose values relative to each other give the mass fractions of the dust grain components, $\kappa_{\nu i}$ is the mass absorption coefficient in cm$^2$ gm$^{-1}$ of dust-grain component *i* at frequency *ν*, and $I_\nu^{PAH}$ is a flux-density template for polycyclic aromatic hydrocarbon (PAH) emission, included for modeling spectra of disks around earlier-type (B-G) young stars. Depending upon the spectral range covered, the East modelers often break the spectra into independent parts, so that as many as six temperature components are used (Bouwman et al. 2008). "Easterners" use five different grain species, in each of three different sizes corresponding to volume-equivalent grain radii of 0.1, 1.5, and 6 μm. The sizes are chosen to represent the full range of sizes for which the dust grains are not so internally optically thick that they cannot produce discrete emission features. However, not all of the sizes in the East recipe have significant weights in all models. The dust species in the current East palette are listed in Table 1.

*West* Used for example by Sargent et al. (2009a, 2009b) and Chen et al. (2006) for analysis of *Spitzer*-IRS spectra. This is also a two-temperature model of the infrared-excess flux density, but each temperature has an optically-thin and an optically-thick component, which would normally each be dominated by emission from different distances from the star:

$$F_\nu = B_\nu(T_c)\left(a_{c,0} + \sum_i a_{c,i}\kappa_{\nu i}\right) + B_\nu(T_w)\left(a_{w,0} + \sum_j a_{w,j}\kappa_{\nu j}\right) \quad . \tag{2}$$

Here the terms mean the same as in Equation (1), except for the subscripts *w* and *c* for warm and cool (typically about 400 K and 120 K). In the following we will refer frequently and interchangeably to these "warm" and "cool" components and to the "inner" and "outer" regions of the disk they represent (radii around 0.6 and 10 AU respectively; Sargent et al. 2009b). The ingredients of the West dust grains are also listed in Table 1. Large crystalline grains are omitted from the West prescription, a choice motivated on both observational and physical grounds. In "Western" experience, large crystalline grains have been included in the prescription before but have never been found to improve the fit to a spectrum. Moreover, large grains are thought to form by accretion of smaller ones, which would result in polycrystalline or amorphous aggregates. Small mineral inclusions would in this case show features as if they had size corresponding to the abundance of the particular mineral constituent within the aggregate (Min et al. 2008a; see also Sargent et al. 2009b).

Table 1. Ingredients of dust-grain models of disk emission

| Species | Opacity reference | Shape and structure* |
|---|---|---|
| *East* | | |
| amorphous $MgFeSiO_4$ | Dorschner et al. 1995 | uniform sphere, 0.1, 1.5, 6 μm |
| amorphous $MgFeSi_2O_6$ | Dorschner et al. 1995 | uniform sphere, 0.1, 1.5, 6 μm |
| forsterite, $Mg_2SiO_4$ | Servoin & Piriou 1973 | DHS, 0.1, 1.5, 6 μm |
| clino enstatite, $MgSiO_3$ | Jäger et al. 1998 | DHS, 0.1, 1.5, 6 μm |
| amorphous $SiO_2$ | Henning & Mutschke 1997 | DHS, 0.1, 1.5, 6 μm |
| *West* | | |
| amorphous $MgFeSiO_4$ | Dorschner et al. 1995 | 60% porous, BrEMT, 5 μm; CDE2, submicron |
| amorphous $MgFeSi_2O_6$ | Jäger et al. 1994 | 60% porous, BrEMT, 5 μm; CDE2, submicron |
| forsterite, $Mg_2SiO_4$ | Sogawa et al. 2006 | tCDE, submicron |
| clino enstatite, $MgSiO_3$ | Chihara et al. 2002 | as measured: ground to submicron size, embedded in KBr |
| annealed $SiO_2$ | Fabian et al. 2000 | |

*DHS = distribution of hollow spheres (Min et al. 2005); BrEMT = Brueggeman effective-medium theory (Bohren and Huffman 1983); CDE2 = weighted continuous distribution of ellipsoids (Bohren and Huffman 1983); tCDE = truncated continuous distribution of ellipsoids (Sargent et al. 2009b).

As solid dust grains grow to dimensions greater than 1-2 μm, internal radiative transfer effects begin measurably to broaden and shift the wavelengths of the mid-infrared silicate absorption features. For solid grains in sizes greater than about 10-20 μm, these effects reduce to near zero the contrast of the silicate features. Thus grain growth can only be studied over a narrow range of sizes with the silicate features. This limitation simplifies the modeling, however, as it reduces substantially the number of components necessary to approximate the actual continuous distribution of grain sizes. To represent these "large grains," the East recipe includes two solid-grain components of volume-equivalent radius 1.5 and 6 μm. The West recipe includes just one, but one chosen by noting that model fits are improved by including porosity (e.g. Voshchinnikov & Henning 2008): 5 µm radius and 60% vacuum, thus corresponding to the internal opacity as solid grains with radius 2 µm, but having scattering cross-section quite different from the smaller solid grains.

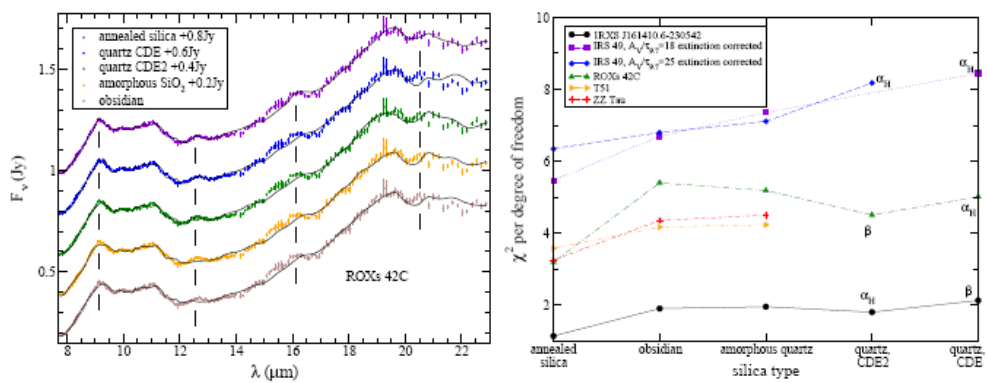

Figure 1: Optimization of silica opacities in the West recipe. Left: best-fitting model spectra compared to that of ROX 42C, in which the only differences in the models are the silica opacities, illustrating that annealed silica (Fabian et al. 2000) fits better than quartz, amorphous or glassy choices. Right: results of silica-opacity variation in models of five objects, showing that annealed silica fits best in each of the objects. From Sargent et al. (2009a).

The two recipes involve different methods for determining the uncertainties of mass fractions, with results that differ systematically between them. Users of the East recipe normally employ a Monte Carlo method, fitting a large number of models to the same spectrum to which artificial Gaussian noise is added that is equivalent to the noise in the original spectrum, and computing the uncertainty in the mass fractions from the dispersion in each obtained in these trials. The uncertainty derived in this fashion is mathematically equivalent to the standard deviation of the mean for purely Gaussian noise. Concerns that most observations are also influenced by systematic effects, such as imperfect flat-fielding, have led users of the West prescription to a much more conservative approach. The uncertainty in a given mass fraction is determined in the West method by the range over which $\chi^2$ per degree of freedom increases by one unit from the minimum value. Thus the uncertainties reported by West users are larger by factors of 5-8 than East, due simply to this difference in convention. The best estimate of uncertainty probably lies somewhere in between: the West method is admittedly too conservative, the East possibly a little too aggressive.

**2.2. Tests of the Tools**

*Identification of the best-fitting interstellar grain analogues.*   The recipes listed in Table 1 are the results of substantial, iterative testing; each group has tried a wide variety of opacities, size ranges, etc. in models of a large selection of protoplanetary disk spectra with high signal-to-noise ratio. Least ambiguous are the choices of the most abundant minerals in small dust grains, each species of which presents several narrow features within the 10 and 20 micron silicate complexes. Let us take the case of silica in the West model as an example. After detailed optimization of the opacities of other silicate species, Sargent et al. (2009a) identified five objects with especially prominent silica features. They proceeded to find the best fits to the spectra varying only the silica opacities, among all the laboratory opacities available for amorphous and crystalline versions. The results are illustrated in Figure 1. A clear winner emerged from this trial: in each case annealed silica (Fabian et al. 2000) produced better-fitting results, by at least one unit of $\chi^2$ per degree of freedom, than quartz, obsidian or amorphous silica. Similar trials and iterations have led both East and West to identify the magnesium silicates forsterite and enstatite as the dominant crystalline silicates in small grains, and to show that all of the small-grain species are quite non-spherical. The East group approximates the departure from spherical shape by assuming a distribution of hollow spheres (DHS; Min et al. 2005). The West uses various forms of a continuous distribution of ellipsoidal shapes (CDE; Bohren and Huffman 1983). The latter approach requires knowledge of the real and imaginary parts of the dielectric constant along all principal optical axes, and reasonable assumptions about the orientation of optical axes with the axes of the ellipsoidal grains (see Sargent et al. 2009a,b).

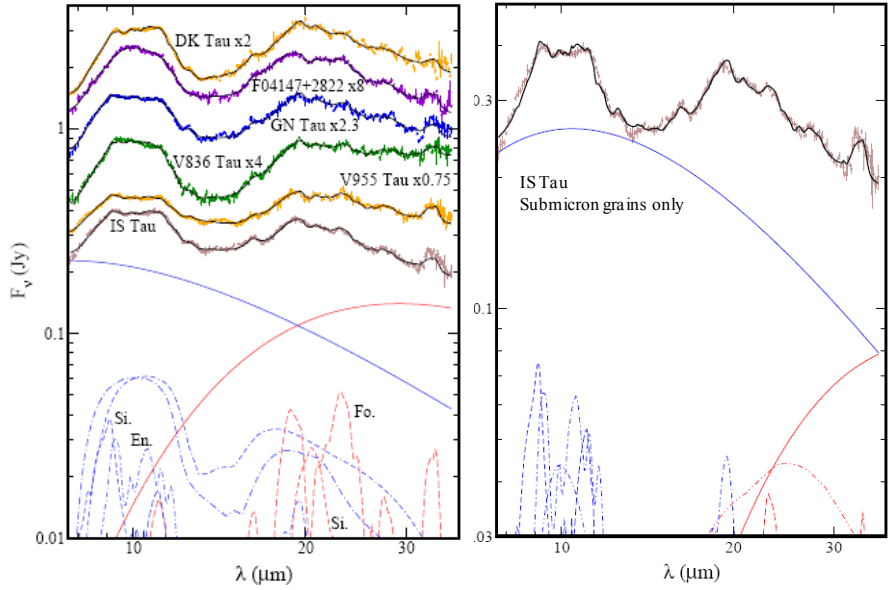

Figure 2: "West" model fits to several Class II YSO spectra. Left: best-fitting models (upper solid curves) compared to observed spectra (points and errorbars); below are the model components for the spectrum of IS Tau: continuous components (solid); submicron forsterite (long dashes), silica (short dashes), and enstatite (dot-short dash); and large amorphous grains of pyroxene (dot-long dash) and olivine (dot-double dash) stoichiometry. Right: best fitting model (continuous curve) of IS Tau (points and errorbars) with only small grains. Again the model components are shown at the bottom of the frame, in the same scheme as at left. This model of IS Tau is visibly inferior to the one with large grains; it is also larger in $\chi^2$ per degree of freedom by 1.6. From Sargent et al. 2009b.

Despite the small size of the parameter space in which they can be detected, large grains are detected in protoplanetary disks. A good example is provided by the spectrum of IS Tau (Figure 2), in which the best-fitting model with large grains is better than the best model that omits large grains, by 1.6 in $\chi^2$ per degree of freedom. Similarly it can be shown that the porosity of such grains must be substantial – 60%, in the West findings – as expected on theoretical grounds (Voshchinnikov & Henning 2008); also, that grain shape makes less difference than it does for the small grains.

*Uniqueness of results* A traditional criticism of the derivation of the mass fraction of dust species is non-uniqueness of the results. This is indeed a danger in silicate spectra of short wavelength span and low signal-to-noise ratio, but the problems are much less with high-quality *ISO* and *Spitzer* spectra. Crystalline mass fractions are usually determined from model fits to multiple narrow features; the mass fractions of large grains constrained by precisely-determined shapes of the smoother underlying emission component. In extreme cases, degeneracy of parameters can be seen in maps of the figure of merit of a fit in the space of concentration of the two degenerate components, as is illustrated in Figure 3 for the two submicron-amorphous grain components in the West recipe. The degeneracies merely indicate that the sums of certain pairs of mass fractions are better determined than either member of the pair. There are dust components – those for which numerous, narrow features are detected, such as forsterite – that are clearly not subject to such degeneracies.

Covariance testing is also used to express such degeneracy statistically, and show in finer detail which derived parameters are subject to degeneracy and which are derived unambiguously. The procedure is described in detail by Sargent et al. (2009b). In the course of minimization of $\chi^2$ in a model fit, one usually calculates a covariance matrix, for which the elements are covariances of pairs of different components, and variances in the parameter values for individual components. From the ratios of covariances and variances one can calculate the correlation coefficient of the covariance, which would approach -1 for perfectly degenerate parameters: that is, a given quality of fit can be achieved with a wide range of the degenerate parameters, reducing one as the other is increased. Similarly, the correlation of the covariance approaches +1 for perfectly non-degenerate parameters, as both are required for a given quality of fit. A study of this sort was carried out recently by Sargent et al. (2009b) on 65 Class II YSOs in the Taurus-Auriga association. There it was found that, although all four amorphous grain species are generally required for good fits, the components of a given grain size with olivine or pyroxene stoichiometry are highly degenerate with each other (correlation of covariance -0.7 to -0.9). Some of the crystalline components that have strong features close to each other are also somewhat degenerate for some temperature regimes, but such degeneracy can usually be broken by comparison of weaker features. For example, submicron enstatite and silica are significantly degenerate in warm material (correlation of covariance -0.5) because their strongest features overlap near $\lambda = 9.2\ \mu\mathrm{m}$. But this degeneracy can be broken because there are other features they do not share, like the isolated silica feature at $\lambda = 12.5\ \mu\mathrm{m}$, that are also not confused with other species. The West recipe's warm submicron forsterite and silica are found in this study always to be highly non-degenerate, as are submicron forsterite and enstatite.

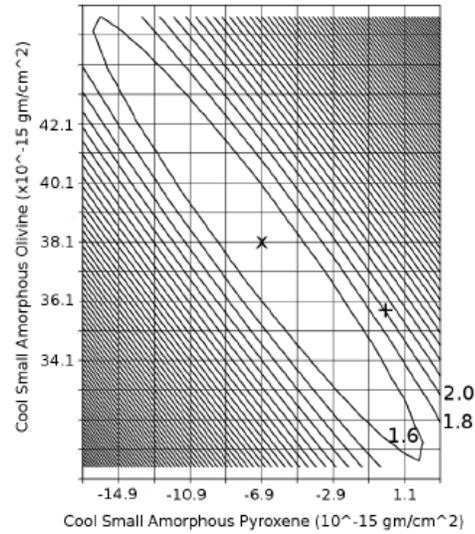

Figure 3: $\chi^2$ per degree of freedom as a function of the two lower-temperature, submicron amorphous dust grain components in a West model fit of the spectrum of the Trapezium $(\theta^1 C\ Ori)$, an exemplar of small amorphous grains. The minimum in $\chi^2$ per degree of freedom is a -45° valley in this space, indicating that the *sum* of the mass fractions submicron amorphous grains of olivine and pyroxene stoichiometry is well determined, but that one component can substitute for the other over a broad range of mass fraction without changing the quality of the model fit. From Sargent et al. (2009b).

*Limitations of plane-parallel, uniform-slab models*    The mid-infrared continuum and broad dust features in protoplanetary disks have been reproduced in accurate detail by models in which the disk structure, heating, cooling, and radiation transport are treated realistically (e.g. D'Alessio et al. 2001, 2006; Dullemond & Dominik 2004, 2008). Trial calculations of an emergent spectrum from a self-consistent disk model with a specified dust mixture – its optical properties included accurately –are followed by modeling with the East or West model. Initial trials of this procedure have been reported by Sargent et al. (2006), Sargent (2008), and Juhász et al. (2009), with the result that the two-temperature models are reasonably accurate, and that the refinements to these models by Juhász (Sicilia-Aguilar et al. 2008, Juhász et al. 2009) do even better, in recovery of the disk-model inputs. Much more characterization of this sort is necessary.

*Spectral Indices for Crystallinity and Grain Size*    Multicomponent dust-feature fitting is difficult enough that many workers in search of trends of grain properties with YSO system properties have often resorted to easily calculated spectral indices designed to serve as proxies of crystalline or large-grain mass fraction (Bouwman et al. 2001; van Boekel et al. 2003; 2005; Przygodda et al. 2003; Kessler-Silacci et al. 2005, 2006, 2007; Watson et al. 2009). The principles behind the indices are that grain growth can appear as a broadening and shift to longer wavelengths of the 10 μm silicate feature, and that the abundant mineral species reliably produce narrow features superposed on the amorphous-silicate features. Usually one expresses the spectral index as the ratio of flux above the continuum at a given wavelength, to that at the peak of a corresponding silicate complex. A widely-used index for studies of grain growth is the "shape" index, based on the fluxes in the 10 μm silicate feature at wavelengths 9.8 and 11.3 μm (e.g van Boekel et al. 2005). A suite of such indices, suitable for tracking warm forsterite, enstatite and silica, and cooler forsterite, is given by Watson et al. (2009); the warm forsterite index is very similar to the "shape" index.

## 2.3. The Dominant Dust Species in Protoplanetary Disks

In application of these models to the spectra of Class II YSOs, good fits ($\chi^2$ per degree of freedom approaching unity) result from models with 8-15 dust components (species and sizes) per temperature component, depending upon the number of features detected. It is unlikely that species we omit are present in quantities similar to those reported for the major species in the recipes. A good example of the state of the art appears in Figure 4. In the following we will describe the grain species that are robustly identified by the models, and list a few species that have been sought but not identified so far, thus summarizing briefly the observations of a few hundred Class II YSOs in nearby associations (Bouwman et al. 2008; Honda et al. 2003, 2006; Sargent et al. 2006, 2009a, 2009b; Sicilia-Aguilar et al. 2007).

*Forsterite* This is the most abundant silicate mineral in protoplanetary disks, detected in virtually all 1-3 Myr-old protoplanetary disks (e.g. 94% in Taurus-Auriga). The mass fraction of forsterite is typically 4-10% but ranges up to about 70%. It is found in all temperature components (thus throughout each disk), all ages < 4 Myr, and throughout the full range of stellar spectral types. There is no sign of any substantial abundance of iron-bearing olivine minerals; studies of the model ingredients always converge on the nearly-pure-magnesium end of the olivine range. Recently there have been indications from models of the very highest signal-to-noise spectra that 10%-iron olivines (that is, crystals with $Mg_{1.8}Fe_{0.2}SiO_4$ stoichiometry) fit slightly, but significantly, better than pure forsterite (Bouwman 2008, private communication). Small forsterite crystals are quite common in primitive solar-system material, with cometary spectra (Crovisier et al. 1997) and cometary samples returned by *Stardust* (e.g. Brownlee et al. 2006) providing ready examples.

*Enstatite* Not much enstatite is seen in the lower temperature components (outer disks) of YSO model fits, but in warmer components it is as ubiquitous as forsterite, and in Taurus we see "warm" mass fractions of 6-7% -- that is, somewhat greater than is typical of forsterite. The monoclinic crystal structure is expected to dominate over orthorhombic, by energetic considerations and by analogy with interplanetary dust particles and meteorites, but at mid-infrared wavelengths the two forms have practically indistinguishable spectra and have been used somewhat indifferently by East and West. As is the case for forsterite, there is no sign of substantial iron content in the pyroxene minerals.

*Silica* In Taurus the mass fraction of silica is typically 2% in the warm or inner-disk components, and 10% in the cooler, outer components. According to the analysis of the West recipe presented above (section 2.2; Figure 1), the closest laboratory analo-gue for the silica in the objects with the most prominent features is the annealed silica characterized by Fabian et al. (2000). This material, in turn, turns out mostly to be composed of the high-temperature polymorphs cristobalite and tridymite, mostly the former. The West recipe usually – though not always – fits the silica features in Class II spectra well. Thus the silica in protoplanetary disks must commonly be in the form of cristobalite and tridymite rather than the lower-temperature polymorphs like α-quartz. This result bears comparison to the "Ada" 81P/Wild sample from *Stardust* (e.g. Zolensky et al. 2006), in which the silica is mostly tridymite. Cristobalite and tridymite are often produced in laboratory smoke-annealing experiments that are meant to approximate conditions in protoplanetary disks (e.g. Rietmeijer et al. 2002).

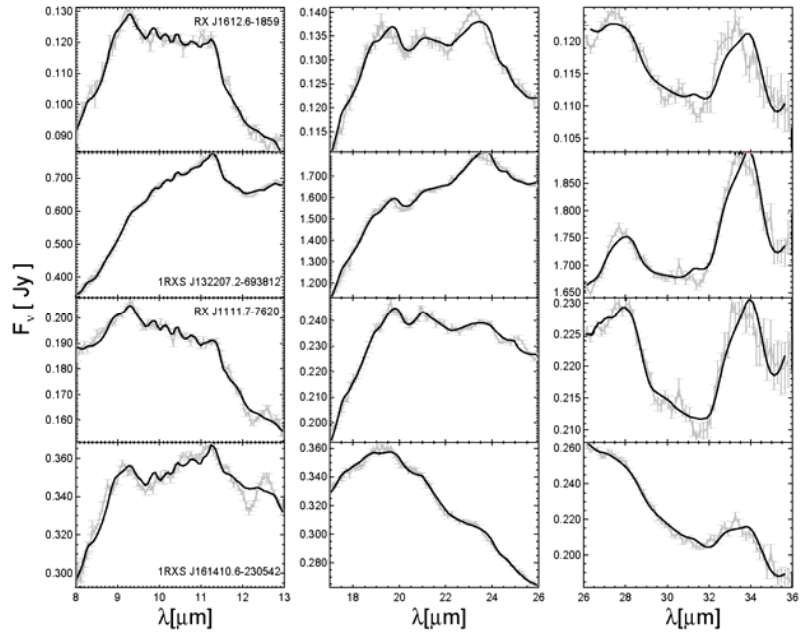

Figure 4: Models of four Class II YSO spectra. For clarity the model spectra are superposed as a dark curve, on top of the lighter data points. The three spectral ranges shown were fit independently, with two-temperature East models. From Bouwman et al. 2008.

*Polycyclic aromatic hydrocarbons (PAHs)*   These features, especially those of the $\lambda = 11-14\,\mu$m complex, appear (often faintly) in most spectra of Class II objects with stellar spectral type about K5 or earlier. PAHs are probably present in all YSOs but lack sufficient excitation to be observed in later-type objects. PAH features in YSOs are discussed elsewhere in this volume, by Gwendolyn Meeus and Luke Keller.

*Ices*   The vast majority of YSOs in which mid-infrared ice features are detected are Class 0/I objects with substantial envelopes. Only when viewed edge-on, do disks in the same evolutionary state as Class II YSOs show features from ices of water, carbon dioxide, ammonia, ammonium and methanol; also the icy material seen in these objects is far from the planet-forming region of the disk. Good examples of ice fea-tures in edge-on disks are CRBR 2422.8-3423 (Pontoppidan et al. 2005) and DG Tau B (Watson et al. 2004).

*Iron-bearing large amorphous silicates* Though the models cannot distinguish sharply between large amorphous grains with olivine or pyroxene stoichiometry, it is nevertheless clear that iron is required in these components for good fits to the spectra. Both East and West recipes use equal parts Mg and Fe in their large amorphous grain components.

*Pristine, interstellar-like grains* A small fraction of protoplanetary disks in the youngest nearby clusters have no signs at all of crystalline silicates or large grains: the mass fraction of amorphous submicron grains is >95% in both inner and outer disk. In Taurus, there are four of these, 5% of the Taurus Class II sample (Watson et al. 2009). Three of the four are the transitional disks CoKu Tau/4, GM Aur, and LkCa 15; the other is UY Aur. Only one of these objects, CoKu Tau/4, lacks dust in its central few´0.1 AU. Otherwise these disks resemble many others in the Taurus association. This is puzzling: it is hard to understand how a 1.5 Myr-old disk with dust at the "sublimation radius" can avoid having detectable crystalline silicates or large grains, given that the time scales for producing substantial mass fractions are < 0.1 Myr. The transitional disks may perhaps be excused because of their unusual radial structure, but complete disks like that in UY Aur cannot.

Not every protoplanetary-disk grain component has been detected that has been anticipated or hoped for. The overwhelming strength of the features of the majority species undoubtedly masks some important signs of minor species. Here is a list of targets still awaiting discovery, or claimed only tentatively.

1. *Iron-bearing minerals* Despite several tantalizing results (e.g. Keller et al. 2002), suggestions of iron-bearing pyroxenes, olivines, troilite, and wustite have not held up.
2. *Hydrated silicates* Phyllosilicates are abundant in some primitive solar-system objects, but not among the small dust grains in protoplanetary disks. Recently Reach et al. (2009) presented what may be the best case so far, tentatively identifying montmorillonite in one of the protoplanetary disks of the IC 1316 association.
3. *Distinctive calcium-aluminum-rich inclusion (CAI) ingredients* CAIs are the oldest components of chondritic meteorites (e.g. Connelly et al. 2008); their appearance is taken to be the time origin for the evolution of the Solar system. Some CAI components have features that lie in a silicate-feature-free part of the mid-infrared spectrum, near $\lambda = 14\ \mu$m (Posch et al. 2007). Hopes that these features would be detected have received a setback, since two frequently-detected, relatively strong spectral features near 14 μm have been identified as rovibrational bands of acetylene and HCN (Carr & Najita 2008).
4. *Alumina, spinel or silicon carbide* Mid-infrared features of these minerals are commonly seen in late-type stellar spectra, but not in protoplanetary disks so far.
5. *Carbonates* The mid-infrared spectrum is not the best place to look for carbonate mineral features, so the lack of reports of carbonates from *Spitzer* spectra is not surprising. *Herschel*, working at far-infrared wavelengths, will have a much better chance.

## 3. Dust Evolution in Class II YSOs

### 3.1. Evolution Over 1-5 Myr

Interstellar dust has no detectable crystalline silicates or large dust grains (Min et al. 2008b), and evidence for such grains in YSO envelopes is slight (see, however, Ciardi et al. 2005); minerals and large grains whose frequent detection is discussed above must mostly have formed *in situ*. But just as is the case for sedimentation of the dust in disks, the formation of minerals and large grains seems to happen quickly: we see large quantities of such processed dust even in the youngest Class II objects. In Taurus-Auriga, for which 10% of the YSOs are Class 0/I protostars, 5% of the Class II objects show no signs of mineralization and grain growth in their mid-infrared spectra (Watson et al. 2009). The median mass fractions of large grains and crystalline grains in the inner parts of sixty-five Taurus disks are 50% and 10% (Sargent et al. 2009b). The corresponding properties of the seven somewhat older protoplanetary disks in the FEPS sample are 92% and 2% (Bouwman et al. 2008). In between we get intermediate median values (Figure 5); the general trend is that small and crystalline grains vanish as time goes on. Even the latter effect is expected from collisional grain coagulation, though there may be other processes too that can amorphotize, rather than simply hide, crystalline grains.

Inner disks tend to have larger fractions of large grains than outer disks, as shown in Figure 6, as expected in all grain-growth models (e.g. Weidenschilling 1997; Dullemond & Dominik 2005), and as seen directly in resolved images of disks around Herbig Ae stars (van Boekel et al. 2004). However, there is no tendency for the crystalline mass fraction to be larger in inner disks. In Taurus (Sargent et al. 2009b), the median mass fraction of enstatite is larger for the inner disk than the outer, but the opposite is true of forsterite and silica, with the result that the median crystalline mass fraction is somewhat larger for outer disks than inner disks.

### 3.2. Evolutionary Trends Within Individual Disk Populations

More striking than the large median mass fractions, and indications of trends with age, are the large dispersions in mass fraction seen even within samples of objects normally considered to have formed within a time less than the median age. Within these dispersions are buried other trends, for which the well-studied Taurus region currently provides the best example. Watson et al. (2009) and Sargent et al. (2009b) calculated the linear correlation coefficients, and the probabilities that these coefficients could result from a random distribution, for all pairs of dust mass fractions and spectral indices, continuum spectral indices, and system properties such as stellar spectral type, accretion rate, and disk mass. Some significant trends revealed thereby – "significant" meaning < 2% probability of having been drawn from a random sample – are listed in Table 2.

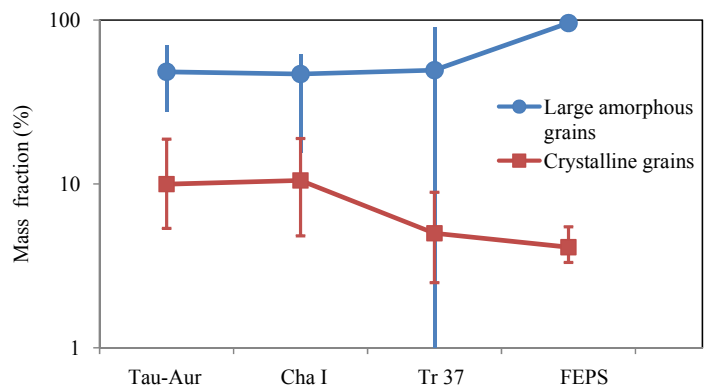

Figure 5: Mass fractions of large ($r$ = 1-10 μm) amorphous grains and crystalline grains in the inner disks (~0.6 AU) of four samples of Class II YSOs. Points and errorbars represent the median and first/third quartiles, respectively, of the mass fractions; the crystalline mass fraction quartile bars are capped, and the others plain. The data come from Sargent et al. (2009b) for the 1.5 Myr-old Taurus Auriga association, unpublished IRS_Disks GTO data for Cha I (2.5 Myr), Sicilia-Aguilar et al. (2007) for the 4 Myr-old Tr 37, and Bouwman et al. (2008) for the FEPS objects, which are thought to be 3-5 Myr of age.

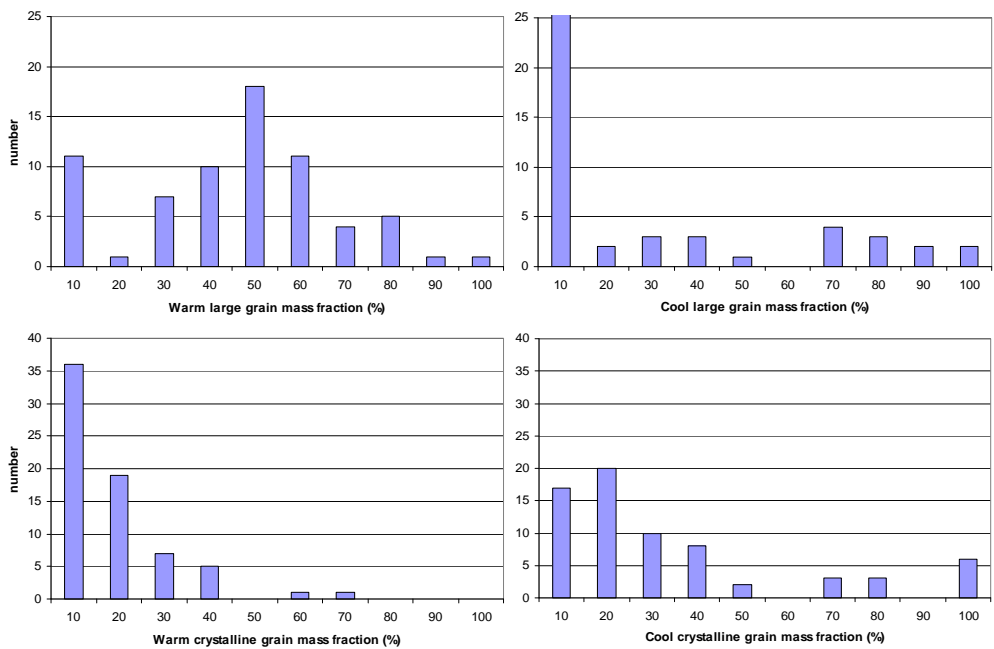

Figure 6: Histograms of large-grain ($r$ = 1-10 μm, top) and crystalline-grain (bottom) mass fraction in inner (~0.6 AU, left) and outer (~10 AU, right) disks in Class II YSOs of the 2 Myr-old Taurus-Auriga association. From Sargent et al. (2009b).

Table 2. Trends involving large or crystalline dust grains (Sargent et al. 2009b)

| Component 1 | Component 2 | Correlation (Pearson's $r$) | $p_{rand}$ (%)* |
|---|---|---|---|
| Warm enstatite | Warm forsterite | 0.62 | 0.0 |
| Warm enstatite | Warm silica | 0.31 | 1.2 |
| Cool forsterite | Cool silica | 0.29 | 2.0 |
| Cool forsterite | Warm large amorphous | 0.32 | 0.9 |
| Cool forsterite | Warm crystalline | 0.32 | 1.0 |
| Warm silica | Cool crystalline | 0.32 | 0.9 |
| Warm silica | Cool forsterite | 0.36 | 0.4 |
| Warm enstatite | Cool forsterite | 0.29 | 1.9 |
| Cool forsterite | $\alpha_{smm}$ § | -0.40 | 2.0 |
| Warm crystalline | $n_{6-13}$ § | -0.50 | 0.0 |
| Warm forsterite | $n_{6-13}$ § | -0.42 | 0.1 |
| Warm enstatite | $n_{6-13}$ § | -0.41 | 0.1 |
| Warm silica | $n_{6-13}$ § | -0.39 | 0.2 |
| Cool enstatite | $n_{13-31}$ § | -0.31 | 1.9 |
| Warm large amorphous | $n_{13-31}$ § | -0.36 | 0.5 |
| Cool large amorphous | $n_{13-31}$ § | -0.32 | 1.5 |
| Warm large amorphous | Stellar multiplicity | 0.33 | 0.7 |
| Warm small amorphous | $M$(star) | 0.40 | 0.1 |
| Warm large amorphous | $M$(star) | -0.36 | 0.5 |

* Probability that the correlation could have resulted from a random distribution.
§ Continuum spectral index; see text for definition.

The strongest trends seen in Table 2 involve the relationships among the dust components themselves, and with the vertical structure of the disk. All pairs of mineral mass fractions are positively correlated. This may indicate that crystallization must not involve much transformation or incorporation of one mineral into another, which would have given negative correlations. The mass fractions of crystalline silicates are negatively correlated with continuum spectral indices, such as the submillimeter slope $\alpha_{smm}$ (Andrews & Williams 2005) and those involving 6, 13 and 31 μm, $n_{6-13}$ and $n_{13-31}$ (Watson et al. 2009). In turn, $n_{6-13}$ and $n_{13-31}$ are indicators of the degree of settling of dust grains to the disk midplane; their values run inversely with the degree of sedimentation (Furlan et al. 2005, 2006, 2009). Large-grain mass fractions are similarly anticorrelated with the mid-infrared spectral indices.

Thus the mass fractions of large grains, and the mass fractions of all three crystalline species, all increase as the degree to which the disks are sedimented increases (Sargent et al. 2009b; Watson et al. 2009). Large grains have long been thought to behave in this way (Goldreich and Ward 1973; Weidenschilling 1997). Few predicions of this sort have been made for the small crystalline grains. Most models have crystalline grains forming very close to the star and being transported outward; only models which would transport the crystals close to disk midplane, like that by Ciesla (2007), would naturally accommodate the trend of crystallinity with sedimentation.

For other properties of YSO systems, there are few trends involving the dust grain composition. Crystallinity is impressively uncorrelated with the masses and temperatures of disks and central stars, and with stellar luminosity and accretion rate (Watson et al. 2009). However, there are three trends between inner-disk large grains and stellar properties. Stellar multiplicity appears to promote large dust grains in the inner disk, possibly from stirring of the disks in multiple systems. Stellar mass is positively correlated with the mass fraction of warm submicron amorphous (i.e. interstellar-like) grains, and negatively correlated with warm large amorphous grains. This is probably an outcome of the tendency for inner disks to have more large grains than outer disks (Figure 6), and the "warm" part of the disk to be further from the star in more luminous systems. Apai et al. (2005) suggested a similar reason for the appearance of the traits of large grains in the spectra of several disks around brown dwarfs.

Despite the high statistical significance of these trends, none of them is particularly tight: there is a large dispersion around each trend. Thus the origin of the large variation of dust properties within a given young association, seen in Figure 5 and Figure 6, is still unidentified. As there is no tight correlation with stellar mass, there is no observed property of disk systems whose range within the sample would naturally give rise to a range of evolutionary time scales, thereby to explain the large dispersions while still involving steady grain-growth or crystallization mechanisms.

### 3.3. No Spectral Index for Grain Growth

Unfortunately there is no index of dust-grain size that works on Class II YSOs less than several million years old, despite common use of the "shape vs. strength" trend of the 10 μm silicate feature (Przygodda et al 2003; Honda et al. 2006; Kessler-Silacci et al. 2005, 2006, 2007; Schegerer et al. 2006). The trend is real, and seems to trace grain growth well in objects with small crystalline mass fractions, such as Herbig Ae stars (van Boekel et al. 2005) and older Class II YSOs (Bouwman et al. 2008). However, in most populations of Class II objects the shape-strength correlation is primarily due to the trend of sedimentation and crystallinity, and can be produced in disks with only small dust grains (Watson et al. 2009). Neither shape nor strength themselves are well correlated with large-grain mass fraction in Class II YSOs. In Taurus, as shown in Figure 7, both "strength" (equivalent width of the 10 μm silicate complex) and "shape" (silicate-feature 11.1/9.8 μm flux ratio, similar to $O_{10}$ in Watson et al. 2009) are strongly linked to crystallinity, but "strength" is not correlated with large-grain mass fraction. The latter result is underscored by the models of individual objects: in the seven objects with the smallest "strengths" in the Taurus association, Sargent et al. (2009b) found in each case that particularly small large-grain mass fractions, and large crystalline-grain mass fractions, were required. The "shape" index is marginally correlated with large-grain mass fraction, but strongly correlated with the forsterite mass fraction, as is also shown in Figure 7.

Spectral indices involving the wavelengths of strong crystalline features, such as those defined by Watson et al. (2009), do indeed function well as proxies of crystalline mass fraction. But grain growth and the trends of large grains will still have to be determined the hard way, by spectral decomposition.

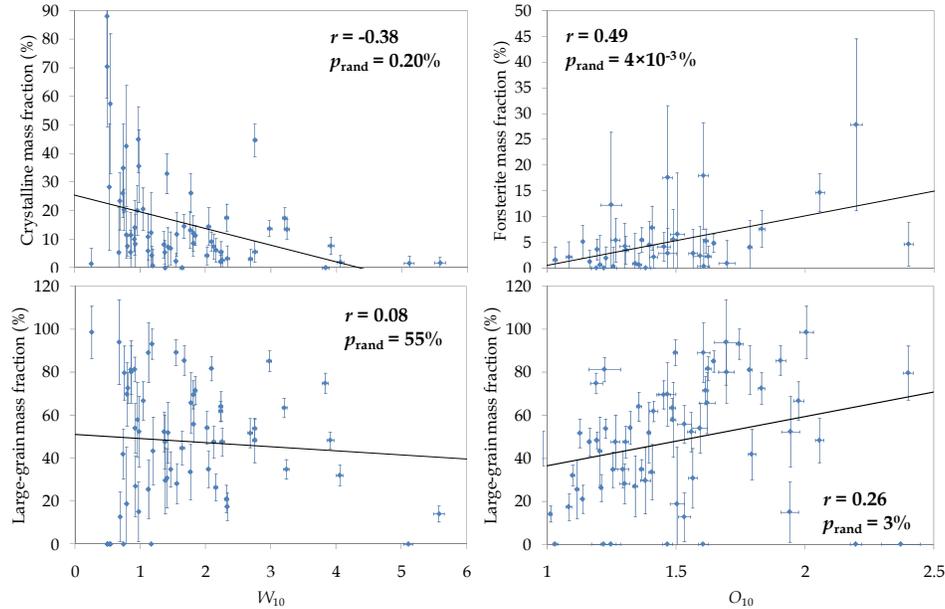

Figure 7: Crystalline and large-grain mass fractions plotted against "strength" ($W_{10}$, equivalent width of the 10 μm silicate complex) and "shape" ($O_{10}$, a narrowband spectral index at 11.1 μm; see Watson et al. 2009) for 65 objects in Taurus (Sargent et al. 2009b). Also plotted are linear correlations, along with the correlation coefficients (Pearson's $r$) and the probabilities that these coefficients could result from a random distribution ($p_{rand}$). Significant correlations in a sample this size have $p_{rand}$ less than about 2%.

### 3.4. Comparison With Dust-processing Models

The standard models of grain growth and crystallization predict that large-grain mass fractions, degree of dust sedimentation to midplane, and crystalline-grain mass fractions increase on time scales much shorter than the ages of Class II YSOs. This would help explain variation in grain properties, if they can make the correct grain components, and if there exist mechanisms for interrupting and/or restarting them.

*Grain growth*   If grains in the silicate emission region grow at rates appropriate for the thermal speeds, they reach sizes of 10 μm or larger in no more than $6 \times 10^4$ years in the outer disk, and faster in the inner disk (Weidenschilling 1997). This is too fast for the large-grain mass fractions found in Taurus (1.5 Myr). Perhaps turbulent motions impede the rate of growth, and inhibit settling of the smaller "large" grains, as suggested by Dullemond & Dominik (2004) and Hubbard & Blackman (2006).

*Crystallization at the dust-sublimation radius*   The models of grain crystallization in widest use involve vaporization of original grains close to star, and re-condensation and crystallization at high temperature and density (Grossman 1972; Gail 2001, 2004). The rate of crystallization depends upon stellar luminosity, but is in any case large: capable of crystallizing the dust in the disk within $10^4$ years. To distribute these crystals throughout the disk, one invokes a radial transport mechanism such as meridional flows (Keller and Gail 2004), X-wind reflux (Shu et al. 2001), or turbulent diffusion (Gail 2001, Ciesla 2007). X-wind reflux provides the crystals from above the disk, which would seem to be inconsistent with a trend of crystallinity with sedimentation. Substantial crystalline mass fractions are found many tens of AU from the star, which is a problem for meridional flows. But turbulent diffusion may work in both respects (Ciesla 2007).

The crystallization models predict abundances close to chemical equilibrium: magnesium silicates, particularly enstatite, and metallic iron should predominate. Indeed there is a tendency for enstatite to be concentrated in the inner disk (see section 2.4). However, enstatite is rarer in outer disks, where forsterite and silica dominate. Thus, along with an efficient radial transport mechanism consistent with the sedimentation trend, one requires conversion of enstatite to forsterite and silica while remaining consistent with the positive correlations among these grain species.

*Crystallization by shock heating* Unlike photo-evaporative mechanisms, shock heating can be episodic, as the related instabilities or drivers come and go. Annealing or melting of amorphous grains in spiral shocks (Harker and Desch 2002) can produce crystallization *in situ*, and produce non-chemical-equilibrium abundances. By this means, forsterite could be made more abundant than enstatite, in outer disks. Shocks may also explain the production of silica, and its tendency to exist in high-temperature polymorphs like cristobalite and tridymite, as observed by Sargent et al. (2009a). Enstatite melts incongruently into forsterite and silica at about 1400 K. When annealed at 1200-1300 K for several hours, silica converts to cristobalite and tridymite (Fabian et al. 2000); subsequent cooling at rates of 10-1000 K per hour would keep the silica from converting to quartz. These peak temperatures and cooling rates are similar to those found in a spiral-shock model by Desch & Connolly (2002) that was designed to produce chondrules. It is possible that this resemblance of the formation conditions of chondrules and high-temperature silica polymorphs indicates a concordance of meteoritic and protoplanetary-disk time scales.

## 4. Conclusions

The study of dust in protoplanetary disks is beginning to contribute to our understanding of disk evolution and the formation of planets. We have found strong correlations of silicate-mineral and large-grain mass fractions with age and degree of disk sedimentation, but with large dispersion about these trends; striking lacks of correlation of mineralization with other stellar and disk properties; and indications that some of the minerals, notably silica, have their origin in processes similar to those thought to give rise to chondrules. These results place strong constraints on the candidates for grain processing and radial transport in disks. They are important in determining the initial conditions of terrestrial-planetary formation, and in aligning the processes and timescales observed in young stellar objects with those of the meteoritic and cometary records of the Solar system. In particular, they are inconsistent with steady application of the standard models for particle growth and mineral formation. Thus either grains are processed by a new mechanisms for which the efficiency can vary widely among systems which appear similar at visible-IR wavelengths, or a conventional method operates, but along with other processes that can erase the trends produced thereby, as discussed by Sargent et al. (2006) and Watson et al. (2009)

There are some 4000 YSOs within 500 pc of the Solar system. By the end of the cryogenic phase of the *Spitzer* mission, the vast majority of them will be observed by *Spitzer*-IRS over its full 5-40 μm range. Only a few hundred of these observations have been digested well so far, and most of those are in the not-necessarily-typical associations within 200 pc. We must keep in mind that samples an order of magnitude larger than those we have just discussed are already in hand, and are likely to change substantially the picture presented above.

**Acknowledgements**.  I am grateful to my *Spitzer*-IRS collaborators, especially Ben Sargent, Bill Forrest, Elise Furlan, Jarron Leisenring, Danielle Nielsen, Cyprian Tayrien and Manoj Puravankara, for permission to discuss many of our results in advance of publication, and to the CDNF organizers, especially Thomas Henning, for the opportunity to present them. The meeting was a very stimulating one and I enjoyed discussions of disk and dust evolution with many colleagues, including instructive ones with Roy van Boekel, Jeroen Bouwman, Kees Dullemond, Attila Juhász, Casey Lisse and Jürgen Steinacker. This work was supported in part by NASA through grants to the *Spitzer*-IRS instrument team, and through the *Spitzer* general-observer program.

**References**

Alexander, R.D., Clarke, C.J., & Pringle, J.E. 2006, MNRAS, 369, 216
Andrews, S.M., & Williams, J.P. 2005, ApJ, 631, 1134
Apai, D., Pascucci, I., Bouwman, J., Natta, A., Henning, Th., & Dullemond, C.P. 2005, Science, 310, 834
Bohren, C.F., & Huffman, D.R. 1983, Absorption and Scattering of Light by Small Particles, New York: Wiley
Bouwman, J., Meeus, G., de Koter, A., Hony, S., Dominik, C., & Waters, L.B.F.M. 2001, A&A, 375, 585
Bouwman, J., Henning, Th., Hillenbrand, L. A., Meyer, M. R., Pascucci, I., et al. 2008, ApJ, 683, 479
Brownlee, D., Tsou, P., Aleon, J., Alexander, C.M.O'D., Araki, T., et al. 2006, Science, 314, 1711
Calvet, N., D'Alessio, P., Watson, D.M., Franco-Hernandez, R.,, Furlan, E., et al. 2005, ApJ, 630, L185
Carr, J.S. & Najita, J.R. 2008, Science, 319, 1504
Castillo-Rogez, J. C., Matson, D. L., Sotin, C.; Johnson, T. V., Lunine, J. I., & Thomas, P. C. 2007, Icarus, 190, 179
Chen, C.H., Sargent, B.A., Bohac, C., Kim, K.H., Leibensperger, E., et al. 2006, ApJS, 166, 351
Chihara, H., Koike, C., Tsuchiyama, A., Tachibana, S., & Sakamoto, D. 2002, A&A, 391, 267
Ciardi, D.R., Telesco, C.M., Packham, C., Gómez-Martin, C., Radomski, J.T., et al. 2005, ApJ, 629, 897
Ciesla, F.J. 2007, Science, 318, 613
Connelly, J.N., Amelin, Y., Krot, A.N., & Bizzarro, M. 2008, ApJ, 675, L121
Crovisier, J., Leech, K., Bockelee-Morvan, D., Brooke, T. Y., Hanner, M. S., et al. 1997, Science, 275, 1904

D'Alessio, P., Calvet, N., & Hartmann, L. 2001, ApJ, 553, 321
D'Alessio, P., Calvet, N., Hartmann, L., Franco-Hernández, R., & Servín, H. 2006, ApJ, 638, 314
Desch, S.J., & Connolly, H.C. 2002, M&PS, 37, 183
Dorschner, J., Begemann, B., Henning, Th., Jaeger, C., & Mutschke, H. 1995, A&A, 300, 503
Dullemond, C.P. & Dominik, C. 2004, A&A, 421, 1075
Dullemond, C.P. & Dominik, C. 2005, A&A, 434, 971
Dullemond, C.P. & Dominik, C. 2008, A&A, 487, 205
Fabian, D., Jäger, C., Henning, Th., Dorschner, J., & Mutschke, H. 2000, A&A, 364, 282
Furlan, E., Calvet, N., D'Alessio, P., Hartmann, L., Forrest, W.J., et al. 2005, ApJ, 628, L65
Furlan, E., Hartmann, L., Calvet, N., D'Alessio, P., Franco-Hernández, R., et al. 2006, ApJS, 165, 568
Furlan, E., Watson, D.M., McClure, M., Manoj, P., Espaillat, C., et al. 2009, ApJ, submitted
Gail, H.-P. 2001, A&A, 378, 192
Gail, H.-P. 2004, A&A, 413, 571
Goldreich, P. & Ward, W.R. 1973, ApJ, 183, 1051
Grossman, L 1972, GeCoA, 36, 597
Haisch, K.E. Jr., Lada, E.A. & Lada, C.J. 2001, ApJ, 553, L153
Harker, D.E. & Desch, S.J. 2002, ApJ, 565, 109
Henning, Th. & Mutschke, H. 1997, A&A, 327, 743
Henning, Th., Il'in, V.B., Krivova, N.A., Michel, B., & Voshchinnikov, N.V. 1999, A&AS, 136, 405
Hernández, J., Hartmann, L., Calvet, N., Jeffries, R.D., Gutermuth, R., et al. 2008, ApJ, 686, 1195
Houck, J. R., Roellig, T. L., van Cleve, J., Forrest, W. J., Herter, T., et al. 2004, ApJS, 154, 18
Honda, M., Kataza, H., Okamoto, Y.K., Miyata, T., Yamashita, T., et al. 2003, ApJ, 585, L59
Honda, M., Kataza, H., Okamoto, Y.K., Yamashita, T., Min, M., et al. 2006, ApJ, 646, 1024
Hubbard, A., & Blackman, E. G. 2006, NewAst, 12, 246
Ida, S. & Lin, D.N.C. 2008, ApJ, 685, 584
Jäger, C., Mutschke, H., Begemann, B., Dorschner, J., & Henning, T. 1994, A&A, 292, 641
Jäger, C., Molster, F.J., Dorschner, J., Henning, Th., Mutschke, H., & Waters, L.B.F.M. 1998, A&A, 339, 904
Jäger, C., Il'in, V.B., Henning, Th. Mutschke, H, Fabian, D., et al. 2003, JQSRT, 79-80, 765
Juhász, A., Henning, Th., Bouwman, J., Dullemond, C.P., Pascucci, I. and Apai, D. 2009, ApJ, in press (arXiv:0902.0405v1)
Kamp, I. &Dullemond, C.P. 2004, ApJ, 615, 991
Keller, Ch. & Gail, H.-P. 2004, A&A, 415, 1177
Keller, L.P., Hony, S., Bradley, J.P., Molster, F.J., Waters, L.B.F.M., et al. 2002, Nature, 417, 148
Kessler-Silacci, J.E., Hillenbrand, L.A., Blake, G.A., & Meyer, M.R. 2005, ApJ, 622, 404
Kessler-Silacci, J., Augereau, J.-C., Dullemond, C.P., Geers, V., Lahuis, F., et al. 2006, ApJ, 639, 275
Kessler-Silacci, J. E., Dullemond, C. P., Augereau, J.-C., Merín, B., Geers, V. C., et

al. 2007, ApJ, 659, 680
Min, M., Hovenier, J.W., & de Koter, A. 2005, A&A, 432, 909
Min, M., Hovenier, J. W., Waters, L.B.F.M., & de Koter, A. 2008a, A&A, 489, 135
Min, M., Waters, L.B.F.M., de Koter, A., Hovenier, J.W., Keller, L.P., & Markwick-Kemper, F. 2008b, A&A, 486, 779
Mordasini, C., Alibert, Y., Benz, W., & Naef, D. 2008, ASPC, 398, 235
Muzerolle, J., Hillenbrand, L., Calvet, N., Briceño, C., & Hartmann, L. 2003, ApJ, 592, 266
Najita, J. & Williams, J. 2005, ApJ, 635, 625
Pontoppidan, K.M., Dullemond, C.P., van Dishoeck, E.F., Blake, G.A., Boogert, A.C.A., et al. 2005, ApJ, 622, 463
Posch, Th., Mutschke, H., Trieloff, M., & Henning, Th. 2007, ApJ, 656, 615
Przygodda, F., van Boekel, R., Àbrahàm, P., Melnikov, S.Y., Waters, L.B.F.M., & Leinert, Ch. 2003, A&A, 412, L43
Reach, W.T., Faied, D., Rho, J., Boogert, A.C.A., Tappe, A., et al. 2009, ApJ, 690, 683
Rietmeijer, F.J.M., Hallenbeck, S.L., Nuth, J.A., & Karner, J.M. 2002, Icarus, 156, 269
Sargent, B., Forrest, W.J., D'Alessio, P., Li, A., Najita, J., et al. 2006, ApJ, 645, 395
Sargent, B.A. 2008, Ph.D. dissertation, University of Rochester
Sargent, B.A., Forrest, W.J., Tayrien, C., McClure, M.K., Li, A., et al. 2009a, ApJ, 690, 1193
Sargent, B.A., Forrest, W.J., Tayrien, C., McClure, M.K., Watson, D.M., et al. 2009b, ApJ, in press (arXiv:0811.3622)
Schegerer, A., Wolf, S., Voshchinnikov, N. V., Przygodda, F., & Kessler-Silacci, J. E. 2006, A&A, 456, 535
Servoin, J.L. & Piriou, B. 1973, PhyStatSol, 55, 677
Shu, F.H., Shang, H., Gounelle, M., Glassgold, A.E., & Lee, T. 2001, ApJ, 548, 1029
Sicilia-Aguilar, A., Hartmann, L., Calvet, N., Megeath, S.T., Muzerolle, J., et al. 2006, ApJ, 638, 897
Sicilia-Aguilar, A., Hartmann, L.W., Watson, D.M., Bohac, C.J., Henning, Th., & Bouwman, J. 2007, ApJ, 659, 1637
Sicilia-Aguilar, A., Henning, Th., Juhász, A., Bouwman, J., Garmire, G., & Garmire, A. 2008, ApJ, 687, 1145
Sogawa, H., Koike, C., Chihara, H., Suto, H., Tachibana, S., et al. 2006, A&A, 451, 357
Van Boekel, R., Min, M., Waters, L.B.F.M., de Koter, A., Dominik, C., et al. 2005, A&A, 437, 189
Van Boekel, R., Min, M., Leinert, Ch., Waters, L.B.F.M., Richichi, A., et al. 2004, Nature, 432, 479
Van Boekel, R., Waters, L.B.F.M., Dominik, C., Bouwman, J., de Koter, A., et al.. 2003, A&A, 400, L21
Voshchinnikov, N.V. & Henning, Th. 2008, A&A, 483, L9
Watson, D.M., Kemper, F., Calvet, N., Keller, L.D., Furlan, E., et al. 2004, ApJS, 154, 391
Watson, D.M., Leisenring, J.M., Furlan, E., Bohac, C.J., Sargent, B., et al. 2009, ApJS, 180, 84.
Weidenschilling, S. J. 1997, Icarus, 127, 290
Zolensky, M.E., Zega, T.J., Yano, H., Wirick, S, Westphal, A.J., et al. 2006, Science, 314, 1735